\documentclass{mn2e}
\usepackage{epsf}
\def\msun{{\rm M_{\odot}}}

\title[Masses, Beaming and Eddington Ratios in ULXs] {Masses, Beaming
  and Eddington Ratios in Ultraluminous X--ray Sources}

\author[A. R. King] {A. R. King$^{1}$ \\
$^1$Theoretical Astrophysics Group, University of Leicester, Leicester
LE1 7RH}

\date{\today}

\volume{000}

\pagerange{\pageref{firstpage}--\pageref{lastpage}} \pubyear{2005}

\begin{document}

\label{firstpage}

\maketitle

\begin{abstract}

I suggest that the beaming factor in bright ULXs varies as $b \propto
\dot m^{-2}$, where $\dot m$ is the Eddington ratio for
accretion. This is required by the observed universal $L_{\rm soft}
\propto T^{-4}$ relation between soft--excess luminosity and
temperature, and is reasonable on general physical grounds. The beam
scaling means that all observable properties of bright ULXs depend
essentially only on the Eddington ratio $\dot m$, and that these
systems vary mainly because the beaming is sensitive to the Eddington
ratio. This suggests that bright ULXs are stellar--mass systems
accreting at Eddington ratios of order 10 -- 30, with beaming factors
$b \ga 0.1$. Lower--luminosity ULXs follow bolometric (not
soft--excess) $L \sim T^4$ correlations and probably represent {\it
  sub}--Eddington accretion on to black holes with masses $\sim
10\msun$. High--mass X--ray binaries containing black holes or neutron
stars and undergoing rapid thermal-- or nuclear--timescale mass
transfer are excellent candidates for explaining both types. If the $b
\propto \dot m^{-2}$ scaling for bright ULXs can be extrapolated to
the Eddington ratios found in SS433, some objects currently identified
as AGN at modest redshifts might actually be ULXs
(`pseudoblazars'). This may explain cases where the active source does
not coincide with the centre of the host galaxy.

\end{abstract}

\begin{keywords}
  accretion, accretion discs -- binaries: close -- X-rays: binaries --
  black hole physics -- galaxies: active -- BL Lac objects
 
\end{keywords}

\section{Introduction}

There are currently two models proposed for ultraluminous X--ray
sources (ULXs). In one they are identified as intermediate--mass black
holes (IMBH) accreting at rates below their Eddington limits.  In the
alternative model, ULXs represent a very bright and unusual phase of
X--ray binary evolution, in which the compact object is fed mass at a
rate $\dot M$ well above the usual Eddington value $\dot M_E$. In the
picture proposed by Shakura \& Sunyaev (1973) radiation pressure
becomes important at the spherization radius $R_{\rm sph} \simeq
27(\dot M/\dot M_E)R_s/4$, where $R_s = 2GM_1/c^2$ is the
Schwarzschild radius of the accreting black hole of mass $M_1$
(Shakura \& Sunyaev, 1973; see also Begelman et al., 2006; Poutanen et
al., 2007). Within this radius the disc remains close to the local
radiation pressure limit. Matter is therefore blown away so that the
accretion rate decreases with disc radius as $\dot M(R) \simeq \dot
M(R/R_{\rm sph}) \simeq \dot M_E(R/R_s)$. As the disc wind has the
local escape velocity at each radius, we see from mass conservation
that the wind is dense near $R_{\rm sph}$ and tenuous near the inner
disc edge, and there is a vacuum funnel along the central disc axis through
which radiation escapes.

In this model the large apparent X--ray luminosity $L_X =
10^{40}L_{40}$~erg\, s$^{-1}$ results from two effects of
super--Eddington accretion (Begelman et al., 2006; Poutanen et al.,
2000). First, the bolometric luminosity is larger than the usual
Eddington limit by a factor $\sim 1 + \ln(\dot M/\dot M_E)$, which can
be of order $5 -10$ for the high mass transfer rates encountered at
various stages of the evolution of a compact stellar--mass
binary. Second, the luminosity of a ULX is collimated by a beaming
factor $b$ via scattering off the walls of the central funnel. (Note
that here and thoughout this paper, `beaming' simply means geometrical
collimation, and not relativistic beaming.)  These conditions could
occur in a state of high mass transfer (cf King, 2001, Rappaport et
al., 2005) or conceivably a transient outburst (King, 2002). In this
picture one would expect on physical grounds that the Eddington ratio
$\dot m = \dot M/\dot M_E$ should determine the beaming factor
$b$. However current modelling has not yet derived this connection,
allowing a spurious extra degree of freedom in comparing this picture
with observations.

A clue here comes from the fact that bright ULXs have spectra
consisting of a power law plus a soft ($kT \sim 0.1 - 0.3$~keV) excess
which can be modelled as a blackbody. This is usually taken as a
multicolour disc with the maximum disc temperature as the reference
value, but the fitted temperature is not very different if the
blackbody is assumed uniform. Feng \& Kaaret (2007) show that the
luminosity $L_{\rm soft}$ and temperature $T$ of the blackbody
component vary as $L_{\rm soft} \propto T^{-n}$ with $n = -3.1\pm 0.5$
in the ultraluminous source NGC 1313 X--2. Kajava \& Poutanen (2008,
hereafter KP)) have recently extended this result to a sample of nine
ULXs (including NGC 1313 X--2) which have a power law continuum with a
soft excess. These are essentially all the sources with inferred
luminosities permanently above $\sim 3\times 10^{39}$~erg\,
s$^{-1}$. Strikingly, KP find that all of these soft--excess objects
cluster around the line
\begin{equation} 
L_{\rm soft} = 7\times 10^{40}T_{0.1{\rm keV}}^{-4}~{\rm erg\, s^{-1}}
\label{KP}
\end{equation}
at all epochs (see the right--hand panel of their Figure 3). Here
$T_{0.1{\rm keV}}$ is the temperature in units of 0.1~keV. KP caution
that the agreement for the coolest and brightest may be affected by an
incorrect subtraction of the hard emission component, but the overall
trend (\ref{KP}) is clear. 

KP also identify a distinct class of `non--power--law' (or thermal)
type ULXs whose {\it medium energy} spectra are fitted by harder
multicolour disc blackbodies with reference temperatures $kT_{\rm
  medium}\sim 0.5 - 2$~keV rather than power laws plus a soft excess.
These systems all have inferred luminosities permanently below $\sim
10^{39}$~erg\, s$^{-1}$. They do not obey (\ref{KP}), but instead
follow individual luminosity -- temperature correlations $L_{\rm
  medium} \propto T_{\rm medium}^4$, just like standard (non--ULX)
black hole binaries (Gierli\'nski \& Done 2004).

At first sight, as KP remark, the correlation (\ref{KP}) for the soft
excesses of bright, power--law ULXs seems counterintuitive, as one might
expect a blackbody to vary as $L_{\rm soft} \propto T^4$. However this
assumes that the characteristic radius $R$ of the blackbody remains
fixed as other parameters vary, and indeed that the inferred $L_{\rm
  soft}$ is not affected by beaming, which could itself also vary
systematically. 

I shall show here that in the picture of ULXs as super--Eddington
accretors suggested by Begelman et al (2006) and Poutanen et
al. (2007), the correlation $L_{\rm soft} \propto T^{-4}$ is actually
expected, and results from a tight relation between the beaming factor
$b$ and the Eddington ratio $\dot m$ of the form $b \sim \dot m^{-2}$.
Using the observed relation (\ref{KP}) we find $b$ and $m_1 =
M_1/\msun$ as functions of the Eddington ratio $\dot m$ for a given
inferred disc luminosity.  With $\dot m$ taking values giving only
modest beaming factors $b \ga 0.1$ we find that the accretors in ULXs
with soft components all have stellar masses $m_1 \la 25$. I suggest
also that the non--power--law ULXs obeying $L_{\rm soft} \propto T^4$
actually have black hole masses sufficiently high ($\sim 10\msun$) that
they are {\it sub}--Eddington.

\section{The $L \sim T^{-4}$ correlation for bright ULXs}

King \& Puchnarewicz (2002) developed a general formalism for treating
blackbody emission from the vicinity of a black hole. They allowed for
geometrical collimation of this emission, but assumed that this did
not change the blackbody luminosity $L$ or temperature $T$. This is
true for example of radiation subject to scattering by
nonrelativistic electrons. King \& Puchnarewicz's main result (their
eqn (5)) is
\begin{equation}
L_{\rm sph} = 2.3\times 10^{44}T_{0.1{\rm keV}}^{-4}{l^2\over pbr^2}
~{\rm erg\, s^{-1}}.
\label{puch}
\end{equation}
Here $L_{\rm sph}$ is the blackbody luminosity an observer would infer
from the observed flux by assuming that it is isotropic, even though
in reality it is collimated by a factor $b$. The quantity $l$ is the
ratio of the intrinsic luminosity to the Eddington limit $L_E$ (which
can exceed unity by the factor $(1 + \ln \dot m)$ mentioned above),
$p$ is a factor allowing for deviations from spherical symmetry in the
source (e.g. that it is actually plane and inclined to the line of sight)
and $r = R/R_s$ is the blackbody radius in units of the Schwarschild
radius.

The derivation of the relation (\ref{puch}) is simple. We express the
intrinsic (pre--collimated) blackbody luminosity as
\begin{equation}
L \propto R^2T^4p \propto M^2T^4r^2p \propto
L^2T^4{r^2p\over l^2},
\end{equation}
where one writes the radius as $R = rR_s \propto rM$ at the first
step, and the mass $M$ as $M \propto L_E \propto Ll^{-1}$ at the
second. Solving this equation for $L$ we find $L \propto T^{-4}$. An
observer assuming that the flux is isotropic with the observed value,
rather than collimated, now infers a total blackbody luminosity $L_{\rm sph}
= b^{-1}L$, i.e.
\begin{equation}
L_{\rm sph} \propto T^{-4}{l^2\over pbr^2},
\end{equation}
which gives (\ref{puch}) when the proportionality constants are included.

King \& Puchnarewicz (2002) used (\ref{puch}) to argue that any source
exceeding the normalization on the rhs must either be super--Eddington
for its mass ($l>1$), or emit from a region much smaller than the
Schwarzschild radius ($r<1$), or emit anisotropically ($pb
<1$). Ultrasoft quasars and some ULXs lie close to this regime on the
$L - T$ plane. Here, setting $L_{\rm sph} = L_{\rm soft}$, we see that
the $ L_{\rm soft} \propto T^{-4}$ correlation (\ref{KP}) for ULX soft
excesses is reproduced provided that
\begin{equation}
{l^2\over pbr^2} = 3\times 10^{-4}.
\label{num}
\end{equation}

Observation thus strongly suggests that $b\propto r^{-2}$. If the
power of $T$ in (\ref{KP}) were not precisely 4, e.g. the value $n =
-3.1 \pm 0.5$ found by Feng \& Kaaret (2007), this would introduce a
$T$--dependence into the relation between $b$ and $r$, i.e. $b \propto
T^{4 - n}r^{-2} \sim T^{0.9}r^{-2}$. Since the fitted value of $T$
varies by a factor $\la 3$, while (as we shall see) the inferred
beaming factor $b$ varies much more, we would make only a small error
in adopting the approximate dependence $b \sim r^{-2}$ here too.

This scaling of $b$ thus seems to be required by
observation. Theoretically, a simple argument suggests that a $b
\propto r^{-2}$ dependence follows from the picture of ULXs proposed
by Begelman et al (2006) and Poutanen et al., (2007), in which a wind
from the accretion disc surface keeps the local accretion rate at the
radiation pressure limit at each disc radius, as originally suggested
by Shakura \& Sunyaev (1973). The outflowing wind is densest near
$R_{\rm sph}$, and has large optical depth both outwards along the
disc plane, and in the vertical direction. Thus most of the disc
radiation emitted within $R_{\rm sph}$ must diffuse inwards by scattering,
until it escapes through the central funnels parallel to the disc
axis. The collimation results from the fact that the funnel is tall
and thin, and has scattering walls.

To apply the formalism leading to (\ref{puch}) we identify the
blackbody radius $R$ as $R \sim R_{\rm sph} = rR_s$, with $r = 27\dot
m\bar r/4$, where $\bar r \sim 1$. The blackbody luminosity emitted by
the disc within $R_{\rm sph}$ is the intrinsic luminosity $L$. This
diffuses inwards and is collimated by the funnels.

For the beaming factor $b$ we consider a simple cylindrical funnel
around the central disc axis. If the typical cylindrical radius of the
funnel is $R_0$, and its height is $H_0 \gg R_0$, the half--angle over
which radiation escapes is $\theta \simeq \sin^{-1} R_0/H_0 \simeq
R_0/H_0$. Then the beaming fraction $b$ is simply the total fractional
area of the two funnels, i.e. $b \sim (1 - \cos\theta) \sim
R_0^2/2H_0^2$. Close to the disc plane, the structure of the central
region of the disc wind (and thus the funnel radius $R_0$) is
independent of the value of Eddington ratio $\dot m >1$ at large $R$,
since all such discs have the same central accretion rate behaviour
$\dot M(R) \simeq \dot M_E(R/R_s)$. We expect that $R_0 \sim \lambda
R_s$ with $\lambda > 1$, as $R_s$ sets the lengthscales in this
region. However the funnel height $H_0$ {\it is} sensitive to $\dot
m$, or equivalently $R_{\rm sph} = rR_s$, as at points far from the
disc plane the wind flow pattern is set by $\dot m$, which is
equivalent to saying that the large--scale flow pattern is
self--similar and scaled by $r$. In particular this requires $H_0 \sim
\mu R_{\rm sph} \propto \mu r$, where $\mu < 1$, so that finally
\begin{equation}
b \simeq {\lambda^2\over 2r^2} \simeq {\lambda^2\over 46\mu^2 \dot m^2}x 
\label{b}
\end{equation}
where $x$ stands for the dimensionless combination 
\begin{equation}
x = {l^2\over p\bar r^2}.
\label{x}
\end{equation}
We see the the observational requirement (\ref{num}) implies
$\lambda/\mu \sim 58$, so that the funnel height is only a few percent
of $R_{\rm sph}$, i.e. $\mu \sim {\rm few} \times 10^{-2}$.  We get
finally
\begin{equation}
b \sim {73\over \dot m^2}x
\label{beam}
\end{equation} 
The reasoning of this paragraph assumes that $\dot m$ is large enough
that the two scales $R_0$ and $R_{\rm sph}$ are very different. The
scaling of $b$ with $\dot m$ is clearly more complex for smaller
$\dot m$. In particular, unless $H_0 > R_0$, which requires $\dot m
> 8.5x^{-1/2}$, one would formally get $b >1$. 

\section{The $L \sim T^4$ correlations for non--power--law ULXs}

The last Section dealt with those ULXs (the majority) for which soft
components are seen, and obey the $L_{\rm soft} \propto T^{-4}$
relation. We noted above that KP show that the remaining
(non--power--law) ULXs follow opposed correlations $L \sim T^4$ for
the medium--energy X--rays. Here the normalization differs for each
individual system. The luminosities of the two groups differ sharply:
the power--law--soft--excess systems have inferred luminosities
permanently above $3\times 10^{39}$~erg\,s$^{-1}$, while the
non--power--law systems are permanently below $10^{39}$~erg\,s$^{-1}$.
It seems likely that these fainter systems correspond to {\it
  sub}--Eddington accretion on to black holes with masses $\ga
10\msun$. It is clear that such systems must exist, and that there is
no reason to expect the collimation leading to the opposite $L_{\rm
  soft} \propto T^{-4}$ behaviour of the bright ULXs. The
  normalizations of the $L \sim T^4$ correlations are then fixed by
the system inclinations and the inner disc radii. The latter are indeed of
order a few Schwarzschild radii for black holes of $\sim 10\msun$.

\section{Masses, Beaming and Eddington Ratios}

We can now check whether the inferred behaviour of the beaming factor
leads to sensible parameters for observed bright ULXs. Although the
relation (\ref{beam}) for $b$ was derived using the inferred blackbody
disc emission, its geometrical nature and the fact that electron
scattering is independent of photon energy makes it probable that it
holds for all forms of ULX luminosity, and indeed even in ULXs where
no blackbody disc component has been identified, provided only that
these correspond to super--Eddington accretion. In particular we can
use the $b \propto \dot m^{-2}$ scaling in considering medium--energy
X--rays, which are generally assumed to carry most of the bolometric
luminosity of a ULX. 

The effect of beaming is to cause an observer to infer a spherical
luminosity
\begin{figure}
\centerline{\epsfxsize9cm \epsfbox{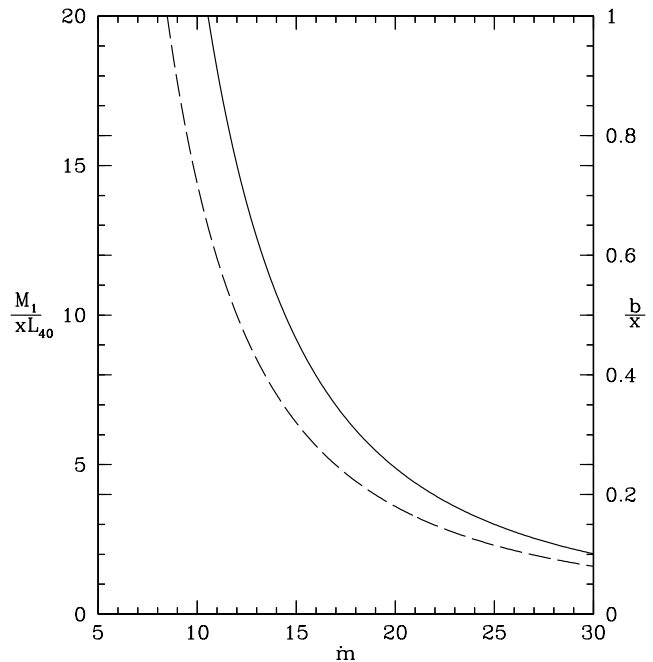}}
\caption{Beaming factor $b$ (dashed curve) and accretor mass $M_1$
  (solid curve, in $\msun$) as functions of the Eddington ratio $\dot m$ for
  ULXs. Here $L_{40}$ is the inferred isotropic bolometric luminosity
  in units of $10^{40}$~erg\, s$^{-1}$ and $x \sim 1$ a dimensionless
  quantity given by equation (\ref{x})}
\label{fig1}
\end{figure}
\begin{equation}
L_{\rm sph} \simeq {L_E\over b}(1 + \ln\dot m)
\label{l}
\end{equation}
(cf Shakura \& Sunyaev, 1973; Begelman et al., 2006).
Eliminating $b$ using (\ref{b}) (or (\ref{num}) gives
\begin{equation}
L_{\rm sph} = 2.2\times 10^{36}m_1\dot m^2(1 + \ln \dot m)x^{-1}~{\rm
  erg\, s^{-1}}.
\label{lum}
\end{equation}
We can re--express this as
\begin{equation}
{m_1\over L_{40}} = {4500\over \dot m^2(1 + \ln \dot m)}x
\label{mass}
\end{equation}
where $L_{40} = L_{\rm sph}/10^{40}$~erg\,s$^{-1}$. 

Figure 1 shows $M_1/L_{40}$ and $b$ as functions of $\dot m$ (using
equations \ref{mass} and \ref{beam}). We see that Eddington ratios in
the range $8.5 < \dot m < 20$ imply stellar masses $1\msun \la m_1 \la
20$ for the accretors if the disc luminosity is $\la 10^{41}$~erg\,
s$^{-1}$, and beaming factors in the range $1 > b \ga 0.2$. Hence
stellar--mass binaries with moderate Eddington ratios and consequently
modest beaming provide very good candidates for explaining ULXs. 

We note from (\ref{lum}) that the inferred luminosity $L_{\rm
  sph}$ varies essentially only because of the sensitivity of the
beaming factor $b$ to $\dot m$. Thus the bolometric luminosity varies
only logarithmically above $L_E$ (assuming that the Eddington ratio
always exceeds unity), but is spread over a smaller or greater solid
angle as $\dot m$ increases or decreases, significantly altering the
inferred luminosity.

Since $\dot M = \dot m\dot M_E \propto (M_1/\eta)\dot m$, where
$\eta$, the radiative efficiency, is similar ($\sim 0.1$) for black
holes and neutron stars, I note that that some ULXs could contain
neutron stars, and could even have lower absolute accretion rates for
the same inferred luminosity. From Fig.~1 we see that a $10\msun$
black hole with $\dot m = 15$ and a neutron star with mass $\la
2\msun$ and $\dot m = 30$ produce similar inferred luminosities, with
the neutron--star system having an absolute accretion rate $\dot M$
lower by a factor $\sim 2.5$ than the black hole. The origin of this
apparently paradox is that the latter system has a smaller beaming
factor. (Put another way, on Fig.~1 the curves of constant $\dot M$
are hyperbolae which cross the hyperbola describing $b$.) For
ultrasoft ULXs with no detectable medium--energy X--ray component,
even white dwarf accretors are possible, particularly since for them
$\eta$ can be enhanced over the pure accretion yield by nuclear
burning of the accreting matter (cf Fabbiano et al., 2003).

\section{ULX populations}

Population studies of ULXs have until now faced the difficulty that
the beaming factor $b$ was not determined, introducing a spurious
degree of freedom. Given the connection (\ref{beam}), we can now
remove this. We consider a population of ULXs with host galaxy space
density $n_g$~Mpc$^{-3}$ and assume that each host contains $N$ ULXs,
with radiation beams oriented randomly. To be in the beam of one such
object one has to search through $\sim 1/Nb$ galaxies, i.e. a space
volume $\sim 1 /n_g N b$. The nearest observed ULX is thus at a
distance
\begin{equation}
D_{\rm min} \sim \left({3\over 4\pi n_g Nb}\right)^{1/3} \sim
0.7(n_gN)^{-1/3}\dot m_1^{2/3}~{\rm Mpc}, 
\end{equation}
where $\dot m_1 = \dot m/10 $. The apparent luminosity of the ULX
is
\begin{equation}
L_{\rm sph} = 2.2 \times 10^{39}m_*\dot m_{1}^2~{\rm erg\, s}^{-1}
\end{equation}
where $m_* = M_1/10\msun$,
giving a maximum apparent bolometric flux 
\begin{equation}
F_{\rm max} = {L_{\rm sph}\over 4\pi D^2} = 4.0 \times
10^{-11}m_*\dot m_{1}^{2/3}(n_gN)^{2/3}~{\rm erg\,
  s^{-1}\, cm^{-2}}
\end{equation}
These relations, together with the results of the previous Section,
agree with the fact that ULXs of apparent luminosity few~$\times
10^{39} - 10^{41}$~erg\, s$^{-1}$ are observed in the Local Group, and
suggest that the typical intrinsic number $N$ per host galaxy is at
most a few. This is in line with estimates of the numbers of
high--mass X--ray binaries in phases of rapid mass transfer on thermal
or nuclear timescales (King et al., 2001; Rappaport et al., 2005),
suggesting that these systems offer good candidates for explaining
most if not all ULXs. Ultimately one needs a population synthesis
calculation to verify that this picture produces the right numbers of
systems with the required moderate Eddington ratios to produce the
nearby ULXs.

\section{Pseudoblazars?}

It is unclear to what value of $\dot m$ one may safely extrapolate the
$b \propto \dot m^{-2}$ dependence inferred here. This is an
interesting question, as we know (cf Begelman et al, 2006; King \&
Begelman, 1999) that the well--studied object SS433 has $\dot m \sim
3000 - 10^4$. Such values are typical for both thermal--timescale and
nuclear--timescale mass transfer from massive donor stars (Rappaport
et al., 2005). 

From the work of the previous Section, now scaling $\dot m$ as $\dot m
= 10^4\dot m_4$, the nearest such object would be at a distance
\begin{equation}
D_{\rm min} \sim \left({3\over 4\pi n_g Nb}\right)^{1/3} \sim
660N^{-1/3}\dot m_4^{2/3}~{\rm Mpc}, 
\end{equation}
where I have taken $n_g \sim 0.02$~Mpc$^{-3}$ as appropriate for $L^*$
galaxies. The apparent isotropic luminosity of such an object would
be
\begin{equation}
L_{\rm sph} = 2.2 \times 10^{45}m_*\dot m_4^2~{\rm erg\, s}^{-1}.
\end{equation}
Hence in distance and apparent luminosity the object would appear as
an AGN. However, unlike a genuine AGN, there is no requirement that it
should lie precisely in the nucleus of the host galaxy.

A possible candidate for such an object is the BL Lac system
PKS~1413+135 (Perlman et al., 2002). With redshift $z = 0.24671$ it
has distance $D \simeq 1000$~Mpc and isotropic luminosity $\simeq
10^{44}~{\rm erg\, s}^{-1}$, but lies at $13 \pm 4$~mas from the
centre of the host galaxy.

\section{Discussion}

The work of this paper suggests that the beaming factor in
super--Eddington accretion varies as $b \propto \dot m^{-2}$. This
seems to be required by observations of the $L_{\rm soft} - T$
correlation, and is reasonable on general geometrical grounds. The
existence of this scaling means that observable properties of ULXs
depend essentially only on the Eddington ratio $\dot m$. If this
conclusion is valid, this removes the spurious degree of freedom
allowing one to choose $b$ independently of $\dot m$ which has made
systematic parameter estimates difficult in the past (e.g. King, 2008,
where these two quantities are not connected).

It appears that most ULXs correspond to stellar mass systems accreting
at Eddington ratios of order 10 -- 30, with corresponding beaming
factors $b \ga 0.1$. High--mass X--ray binaries containing black holes
or neutron stars are excellent candidates, although population
synthesis studies are needed to check this. The scaling inferred here
suggests that ULXs vary mainly because the beaming factor is sensitive
to the Eddington ratio. If the scaling can be extrapolated to the
Eddington ratios found in SS433, some objects currently identified as
AGN at modest redshifts might actually be ULXs. This may explain cases
where the AGN does not coincide with the centre of the host galaxy.

\section{Acknowledgments}
I thank Paul O'Brien, Gordon Stewart and Graham Wynn for valuable
discussions, and the referee Jean--Pierre Lasota for a very helpful
report.


\begin{thebibliography}{}

\bibitem{}

\bibitem[\protect\citeauthoryear{Begelman, King, \& 
Pringle}{2006}]{2006MNRAS.370..399B} Begelman M.~C., King A.~R., Pringle 
J.~E., 2006, MNRAS, 370, 399 

\bibitem[\protect\citeauthoryear{Fabbiano et 
al.}{2003}]{2003ApJ...591..843F} Fabbiano G., King A.~R., Zezas A., Ponman 
T.~J., Rots A., Schweizer F., 2003, ApJ, 591, 843 


\bibitem[\protect\citeauthoryear{Feng 
\& Kaaret}{2007}]{2007ApJ...660L.113F} Feng H., Kaaret P., 2007, ApJ, 660, L113 


\bibitem[\protect\citeauthoryear{Gierli{\'n}ski \&
    Done}{2004}]{2004MNRAS.347..885G} Gierli{\'n}ski M., Done C.,
  2004, MNRAS, 347, 885


\bibitem[\protect\citeauthoryear{Kajava \&
    Poutanen}{2008}]{2008arXiv0809.4634K} Kajava J.~J.~E., Poutanen
  J., 2008, arXiv, arXiv:0809.4634


\bibitem[\protect\citeauthoryear{King}{2002}]{2002MNRAS.335L..13K}
  King A.~R., 2002, MNRAS, 335, L13

\bibitem[\protect\citeauthoryear{King}{2008}]{2008MNRAS.385L.113K} King 
A.~R., 2008, MNRAS, 385, L113 


\bibitem[\protect\citeauthoryear{King \& 
Begelman}{1999}]{1999ApJ...519L.169K} King A.~R., Begelman M.~C., 1999, 
ApJ, 519, L169 


\bibitem[\protect\citeauthoryear{King et
    al.}{2001}]{2001ApJ...552L.109K} King A.~R., Davies M.~B., Ward
  M.~J., Fabbiano G., Elvis M., 2001, ApJ, 552, L109


\bibitem[\protect\citeauthoryear{King \&
    Puchnarewicz}{2002}]{2002MNRAS.336..445K} King A.~R., Puchnarewicz
  E.~M., 2002, MNRAS, 336, 445



\bibitem[\protect\citeauthoryear{Perlman et 
al.}{2002}]{2002AJ....124.2401P} Perlman E.~S., Stocke J.~T., Carilli 
C.~L., Sugiho M., Tashiro M., Madejski G., Wang Q.~D., Conway J., 2002, AJ, 
124, 2401 


\bibitem[\protect\citeauthoryear{Poutanen et 
al.}{2007}]{2007MNRAS.377.1187P} Poutanen J., Lipunova G., Fabrika S., 
Butkevich A.~G., Abolmasov P., 2007, MNRAS, 377, 1187

\bibitem[\protect\citeauthoryear{Rappaport, Podsiadlowski, \&
    Pfahl}{2005}]{2005MNRAS.356..401R} Rappaport S.~A., Podsiadlowski
  P., Pfahl E., 2005, MNRAS, 356, 401



\bibitem[\protect\citeauthoryear{Shakura \& 
Sunyaev}{1973}]{1973A&A....24..337S} Shakura N.~I., Sunyaev R.~A., 1973, 
A\&A, 24, 337 

\end{thebibliography}
\end{document}